\documentclass[preprint,showpacs,preprintnumbers,amsmath,amssymb]{revtex4}

\usepackage{graphicx}
\usepackage{dcolumn}
\usepackage{bm}

\begin{document}

\title{Particle-like property of vacuum states}

\author{Wenzhuo Zhang}
\email{stzwz@siom.ac.cn} \affiliation{Key Laboratory of Quantum
Optics, Shanghai Institute of Optics and fine Mechanics, Chinese
Academy of Sciences. Shanghai, P. R. China, 201800.}
\date{\today}

\begin{abstract}
The wave-particle duality of the vacuum states of quantum fields is
considered and the particle-like property of the vacuum state of a
quantum field is proposed as a vacuum-particle which carries the
vacuum-energy and the vacuum-momentum of this field. The
vacuum-particles can be introduced into the quantum field theory
(QFT) naturally without disturbing its mathematical structures and
calculation results, but makes the QFT more self-consistent. For
instance, the interactions between charged particles and vacuum
state of electromagnetic field appears automatically in the Feynman
diagrams of the quantum electrodynamics (QED), with which the atomic
spontaneous emission and the Casimir effect can be interpreted
directly by the QED perturbation theory. Besides, the relation
between vacuum-particles and spontaneous symmetry breaking is also
discussed.
\end{abstract}


\maketitle

\section{Introduction}

Wave-particle duality is the central concept of quantum mechanics.
It originated from the study of both wave-like and particle-like
nature of lights, and was generalized to matters by Louis de Broglie
in 1924 \cite{deBroglie}. Later the de Broglie hypothesis, which
argues that all matter and energy exhibits both wave-like and
particle-like properties, was left to Erwin Schrodinger to discover
the wave equations of quantum mechanics in 1926. Up to the present
time many experiments have proved that all known particles have the
wave-like property. However, the vacuum states of quantized fields,
which carry non-zero energy, has not been treat by wave-particle
duality yet.

An example is the Klein-Gordon field, whose momentum and energy
after canonical quantization are showed in Eq.\ref{one}.
\begin{equation}\label{one}
\begin{split}
\mathbf{P}=\sum_k\big(a^{+}_ka_k+\frac{1}{2}\big)\mathbf{p}_k\\
H=\sum_k\big(a^+_ka_k+\frac{1}{2}\big)\omega_k
\end{split}
\end{equation}
Here $\omega_k=\sqrt{\mathbf{p}^2+m^2}$ is the frequency of the mode
with wave-vector $\mathbf{k}$, where $k$ is the four-momentum. We
adopt the natural unit $\hbar=c=1$ in the whole paper. In Eq.(1),
when the eigenvalues of the particle-number operators in every mode
($a^+_k, a_k$) are equal to zero, the Klein-Gordon field has the
vacuum momentum (the sum of all modes is zero)
$\sum_k\frac{\mathbf{p}_k}{2}$ and the vacuum energy
$\sum_k\frac{\omega_k}{2}$.

The vacuum state of the Klein-Gordon field are treated as the plane
waves of different modes ($\frac{\mathbf{p}_k}{2}$,
$\frac{\omega_k}{2}$). An interesting question is what may occur if
the vacuum state has the same wave-particle duality to all
particles? It means that the vacuum states may also have the
particle-like property like the excited states of fields
(particles). For the vacuum state of the Klein-Gordon field, a plane
wave with ($\frac{\mathbf{p}_k}{2}$, $\frac{\omega_k}{2}$) is
mathematically equivalent to a vacuum-particle who has the momentum
$\frac{\mathbf{p}_k}{2}$ and the energy $\frac{\omega_k}{2}$. The
vacuum-particle can be excited to an one-particle state with
($\frac{3\mathbf{p}_k}{2}$, $\frac{3\omega_k}{2}$) by absorbing
momentum $\frac{\mathbf{p}_k}{2}$ and energy $\frac{\omega_k}{2}$ (a
particle of the Klein-Gordon field).

So we are interested in study such possibility. The paper is
arranged as follows: In section 2, a vacuum particle hypothesis
(VPH) is presented and the new lines which denote to
vacuum-particles are introduced into Feynman diagrams automatically.
In section 3, we apply the VPH to the QED perturbation theory
without disturbing the calculation of all S-matrix elements, but
exhibit the role of vacuum-particles of dirac field and
vacuum-photons in QED, with which the Casimir effect and the atomic
spontaneous emission is studied by the QED perturbation theory more
reasonably. Section 4 contents the discussion of the relationship
between vacuum-particles and the spontaneous symmetry breaking, as
well as the conclusion.

\section{Vacuum-Particle Hypothesis (VPH)}

A field is always defined and quantized at every point of the space,
and the vacuum energy or the momentum of a field is the integrating
value all over the space. We have proposed that the wave-particle
duality of vacuum states is valid in the introduction, then the
vacuum energy and momentum of a single-mode field are carried by a
particle-like object named "vacuum-particle". We present a
Vacuum-Particle Hypothesis (VPH) with two postulates:

1. Our space is filled with vacuum particles, each of which is the
vacuum state of a single-mode quantum field and it carries the
vacuum-momentum and vacuum-energy of this field.

2. The $N$-particle state of a single-mode field is that the
vacuum-particle of this field overlapped by $N$ field quanta
(particles) of this field.

The two postulates are easy to understand by analogizing with a
quantum harmonic oscillator. Its lowest energy state
$\frac{1}{2}\hbar\omega$ is analogized to the energy a
vacuum-particle, its $N$ quanta excited state
$(N+\frac{1}{2})\hbar\omega$ is analogized to the N-particle state
of a single-mode field, and its quantum $\hbar\omega$ is analogized
to a field quantum (particle). Through the whole paper, the word
"particle" always denotes to a field quantum and the word
"$N$-particle state" always denotes to the state that a vacuum state
overlapped by $N$ field quanta.

According to postulate 1, all vacuum-particles need to be on-shell
because the vacuum state is one of the eigenstate of a quantum field
and the vacuum energy and momentum are always on the mass shell to
compose the four-momentum of a vacuum-particle. According to
postulate 2, the particle and the vacuum-particle of a single-mode
field must have the same velocity so they can overlapped each other
all the time. For a massive field, it means the momentum difference
between the vacuum-particle and the particle comes from the
difference of their proper mass. For a massless field, the
same-velocity condition is automatically obeyed.

\begin{figure}
\includegraphics[width=2.in]{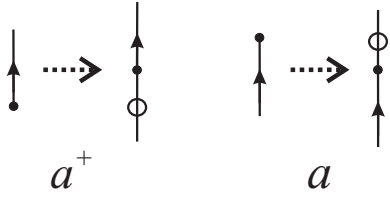}
\caption{Simplified Feynman diagrams of the creation and
annihilation processes of a Klein-Gordon particle. Solid lines with
arrow denote to one-particle state of Klein-Gordon field, Solid
lines with a circle denote to a vacuum-particle of Klein-Gordon
field. The dot lines with arrows show the transformation of Feynman
diagrams when the lines of vacuum particles are introduced.}
\end{figure}

A vacuum particle can be naturally introduced into the physical
processes of creation and annihilation of a particle in the quantum
field theory (QFT). Figure 1 shows the difference between the
Feynman diagrams of creating/annihilating a particle of Klein-Gordon
field before and after the vacuum particle of the Klein-Gordon field
is introduced. One is for the creation process
\begin{equation}
a_k^+|\cdots,0_k,\cdots\rangle=|\cdots,1_k,\cdots\rangle,
\end{equation}
and the other is for the annihilation process
\begin{equation}
a_k|\cdots,1_k,\cdots\rangle=|\cdots,0_k,\cdots\rangle.
\end{equation}
Here both the creation and the annihilation operator have the new
physical meanings. For the Klein-Gordon field, the creation operator
$a_k^+$ in the mode $k$ means exciting a vacuum-particle with
four-momentum $\frac{k}{2}$ into the one-particle state with
four-momentum $\frac{3k}{2}$, while the four-momentum increase
($\frac{3k}{2}-\frac{k}{2}$) equals to the four-momentum of the
field quantum $k$. The annihilation operator $a_k$ in the $k$ mode
means transferring an one-particle state with four-momentum
$\frac{3k}{2}$ into the vacuum-particle with four-momentum
$\frac{k}{2}$, while the four-momentum loss is
($\frac{3k}{2}-\frac{k}{2}=k$). The field operator is
\begin{equation}
\phi(x)=\frac{1}{(2\pi)^{3/2}}\sum_k\sqrt{\frac{1}{2\omega_k}}(a_k^+e^{-ikx}+a_ke^{ikx}).
\end{equation}
Here the four momentum of the field quantum $k$ has the new physical
meaning of $k=\frac{3k}{2}-\frac{k}{2}$.

With the discussion above, we see after the VPH is introduced, only
the energy of every eigenstate of a quantum field is redefined. The
field quantum, field operator as well as the Lagrangian are all
unchanged. Therefore the introduction of vacuum-particles does not
disturb any calculation process of quantum field theory (QFT).
Instead, the vacuum states of quantum fields acquire more clear
physical meanings and recover their momentum and energy by the VPH
in the physical processes of QFT. It makes the QFT more
self-consistent than removing the vacuum-momentum and vacuum-energy
of a field by hand (the trick which is always used in the QFT). In
next section we will focus on the role that the vacuum-particles of
the Dirac field and the electromagnetic field (vacuum-photons) plays
in the basic processes of QED perturbation theory.

\section{quantum electrodynamics with the VPH}

The interaction between charged particles and photons is well
described by the quantum electrodynamics (QED). The calculation
results of QED has showed great agreements with the experimental
data. However, there are some phenomena that caused by the vacuum
states of electromagnetic fields are widely studied, but has been
not treated by QED yet. One is the Casimir effect \cite{casimir},
which has been observed and measured in the past few years.
\cite{measure1,measure2,measure3}. Another is the atomic spontaneous
emission, which has been explained by the interactions between the
excited electrons of atoms and the vacuum states of the
electromagnetic fields. It can be controlled by cavities
\cite{cavity1,cavity2,cavity3}. The two phenomena imply that the
vacuum states of the electromagnetic fields can interact with the
electric charged particles, but they are not described directly by
the QED since the vacuum energy and momentum of the electromagnetic
fields are replaced by hand in the QED.

In this section, we will merge the VPH into the QED perturbation
theory naturally by redefining the momentum and the energy of every
eigenstate, and the physical meaning of the creation and
annihilation operators of Dirac and electromagnetic field (as we did
to Klein-Gordon field in section 2). This process does not disturb
the mathematical form of QED but can describe the coupling of
electric charged particles, photons and vacuum particles of the two
fields, with which the atomic spontaneous emission and the Casimir
effect can be interpreted and studied directly by the QED
perturbation theory.

Vacuum energy and momentum of the Dirac field are showed in
Eq.\ref{two}.
\begin{equation}\label{two}
\begin{split}
&\mathbf{P}_{Dirac}=\sum_{p}\sum_{\sigma=1}^2\big(c_{\sigma
p}^{+}c_{\sigma p} +d_{\sigma p}^{+}d_{\sigma p}-1\big)\mathbf{p}\\
&H_{Dirac}=\sum_p\sum_{\sigma=1}^2\big(c_{\sigma p}^{+}c_{\sigma p}
+d_{\sigma p}^{+}d_{\sigma p}-1\big)E_p
\end{split}
\end{equation}
Here $\mathbf{P}$ is the momentum vector and $p$ is the
four-momentum. When the eigenvalues of the particle-number operator
of electrons ($c_{\sigma p}^{+}c_{\sigma p}$) and positrons
($d_{\sigma p}^{+}d_{\sigma p}$) are all equal to zero in every
mode, according to the VPH, the vacuum state of a Dirac field is a
vacuum-particle which can be excited to either an one-electron state
or an one-positron state. Every vacuum-particle of the Dirac field
has a negative momentum eigenvalue $-\mathbf{p}$ and a negative
energy eigenvalue $-E_p=-\sqrt{\mathbf{p}^2+m^2}$, which are just
the additive inverses of the momentum and energy of a field quantum
(either an electron or a positron) with four-momentum $p$. It is
still a problem that the vacuum energy of Dirac field is negative.
Although some works have tried to solve it, this problem is beyond
the aim of this paper so we treat them as negative values in the
whole paper.

Similar as Eq.\ref{two}, vacuum energy and momentum of the
electromagnetic field can be found in Eq.\ref{three} when the
eigenvalue of the photon-number operator $a_{\lambda
k}^{*}a_{\lambda k}$ equals to zero in every mode. Eq.\ref{three}
are the results of the quantization of electromagnetic field in the
Lorentz gauge, where $\mathbf{k}$ is the momentum vector and $k$ is
the four-momentum.
\begin{equation}\label{three}
\begin{split}
&\mathbf{P}_{EM}=\sum_k\sum_{\lambda=1}^4\big(a_{\lambda
k}^{*}a_{\lambda k}+\frac{1}{2}\big)\mathbf{k}\\
&H_{EM}=\sum_k\sum_{\lambda=1}^4\big(a_{\lambda k}^{*}a_{\lambda
k}+\frac{1}{2}\big)\omega_k,
\end{split}
\end{equation}
Here $a_{\lambda k}^{*}=a_{\lambda k}^{+}$ when $\lambda=1,2,3$ and
$a_{\lambda k}^{*}=-a_{\lambda k}^{+}$ when $\lambda=4$.
$\lambda=1,2$ means the two transverse polarizations, $\lambda=3$
means the longitudinal polarization, and $\lambda=4$ means the
time-like polarization. According to the VPH, a vacuum-particle of
the electromagnetic field, which is a vacuum-photon, has half energy
$\frac{\omega_k}{2}$ and half momentum $\frac{\mathbf{k}}{2}$ of a
quantum of electromagnetic field (a photon) with four-momentum $k$,
while the speed of all vacuum-photons are always \emph{c}.

After the VPH is applied to the vacuum states of the Dirac field and
the electromagnetic field, we can give the new physical meanings of
their creation and annihilation operators as we did to the
Klein-Gordon field in section 2. For the Dirac field,
$c^+_p$/$d^+_p$ means exciting a vacuum-particle with four-momentum
$-p$ into the one-electron/positron state with four-momentum $0$,
while the annihilation operator $c_p$/$d_p$ means the opposite
process. The four-momentum increase or loss is $0-(-p)=p$, which
equals to a Dirac field quanta $p$. For the electromagnetic field,
the creation operator $a_k^+$ means exciting a vacuum-photon with
four-momentum $\frac{k}{2}$ (or a N-photon state with four-momentum
$(N+\frac{1}{2})k$) into the one-photon state with four-momentum
$\frac{3k}{2}$ or a (N+1)-photon state with four-momentum
$(N+\frac{3}{2})k$), and the annihilation operator $a_k$ means the
opposite process. The four-momentum increase or loss
$\frac{3k}{2}-\frac{k}{2}=k$ equals to the field quanta (a photon)
with four-momentum $k$.

\begin{figure}
\includegraphics[width=3.6in]{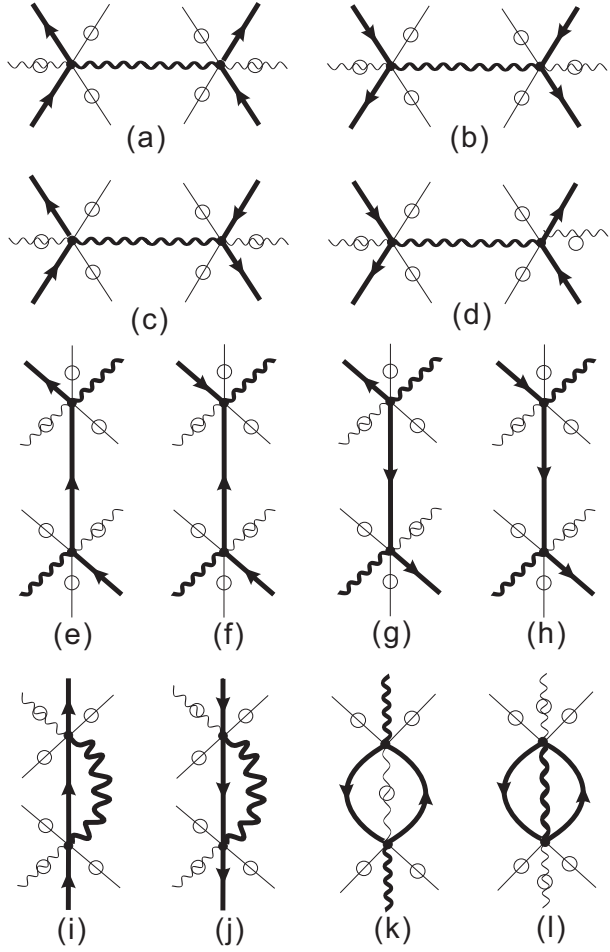}
\caption{Feynman diagrams of every second-order S-matrix element of
the QED perturbation theory. Lines with up-arrow denote to electrons
overlapped by their vacuum-particles and lines with down-arrow
denote to positrons overlapped by their vacuum-particles. Lines with
a circle denote to vacuum-particles of Dirac fields. Wave lines
denote to photons overlapped by their vacuum-photons and wave lines
with circles denote to vacuum-photons.}
\end{figure}

Now the VPH can be mounted to QED perturbation theory. The field
quanta and field operators of the Dirac field and the
electromagnetic field are unchanged. The Lagrangian of QED is also
unchanged. Only the vacuum state of the Dirac field is redefined as
vacuum-particles with four-momentum ${\sum_p}-p$, and the vacuum
state of the electromagnetic field is redefined as vacuum-photons
with four-momentum ${\sum_k}\frac{k}{2}$. With the new physical
meanings of the creation and annihilation operators of the Dirac
field and the electromagnetic field, the Feynman diagrams of QED
need to change into the new forms, where all three-line vertexes
need to be replaced by six-line vertexes. It is just like we did to
the Klein-Gordon field in figure 1. In the QED perturbation theory,
all physical process according to the first-order S-matrix elements
can not obey the energy and momentum conservation at same time, so
the most basic physical process are described by second-order
S-matrix elements, which are
\begin{equation}
\begin{split}
S_{fi}^{(2)}&=\langle
f|-ie\big[\widetilde{\varphi}(x_1)A_{\mu}(x_1)\underline{\varphi(x_1)\widetilde{\varphi}(x_2)}A_{\mu}(x_2)\varphi(x_2)\\
&+\widetilde{\varphi}(x_1)\underline{A_{\mu}(x_1)\varphi(x_1)\widetilde{\varphi}(x_2)A_{\mu}(x_2)}\varphi(x_2)\\
&+\underline{\widetilde{\varphi}(x_1)A_{\mu}(x_1)\varphi(x_1)\widetilde{\varphi}(x_2)A_{\mu}(x_2)\varphi(x_2)}\\
&+\widetilde{\varphi}(x_1)\underline{A_{\mu}(x_1)\underline{\varphi(x_1)\widetilde{\varphi}(x_2)}A_{\mu}(x_2)}\varphi(x_2)\\
&+\underline{\widetilde{\varphi}(x_1)\underline{A_{\mu}(x_1)\varphi(x_1)\widetilde{\varphi}(x_2)A_{\mu}(x_2)}\varphi(x_2)}\\
&+\underline{\widetilde{\varphi}(x_1)A_{\mu}(x_1)\underline{\varphi(x_1)\widetilde{\varphi}(x_2)}A_{\mu}(x_2)\varphi(x_2)}\\
&+\underline{\widetilde{\varphi}(x_1)\underline{A_{\mu}(x_1)\underline{\varphi(x_1)\widetilde{\varphi}(x_2)}A_{\mu}(x_2)}\varphi(x_2)}\big]|i\rangle,
\end{split}
\end{equation}
where $\varphi(x)$ is the field operator of Dirac field and
$A_{\mu}(x)$ is the field operator of electromagnetic field.

Figure 2 shows the new Feynman diagrams corresponding the
Second-order S-matrix elements. The lines of vacuum-photons emerges
in the Feynman diagrams of QED perturbation theory. Phenomena cause
by the interaction between vacuum-states of electromagnetic fields
(vacuum-photons) and charged particles can be studied directly by
the QED perturbation theory now. We see the vacuum-photons
participle every physical process in figure 2 as the external-lines,
whose four-momentum $\frac{k}{2}$ is invariant from initial state
$|i\rangle$ to final state $\langle f|$. However, in some physical
process corresponding to higher S-matrix elements, the four-momentum
of a vacuum-photon may be changed from the initial state to the
final state at a small space-time region, which can be considered as
an interpret of Casimir effect from the QED view. For example,
figure 3 is a case corresponding to a fourth-order S-matrix element.
If the distances between each two of the space-time points $x_1$,
$x_2$, $x_3$, and $x_4$ fit the condition
\begin{equation}
|x_1-x_3|\approx|x_2-x_4|\gg|x_1-x_2|\approx|x_3-x_4|,
\end{equation}
the process can be considered as a vacuum-photon with four-momentum
$k_0$ is absorbed at $x_1$ and is emitted with four-momentum $k_0'$
at $x_2$, which is just a scattering process of a vacuum-photon near
the space-time points $x_1$ and $x_2$. The Casimir effect
\cite{casimir} requires such scattering processes to describe how a
vacuum-photon can be reflected by the surface of a conducting plate,
then the radiation force from the vacuum-photon can act on the
surface of the conducting plate. According to the wave-particle
duality, this process is equivalent to it that the plate reflects
the plane wave of the vacuum-state of the electromagnetic field,
which is the original interpret of the Casimir force \cite{casimir}.

\begin{figure}
\includegraphics[width=2.in]{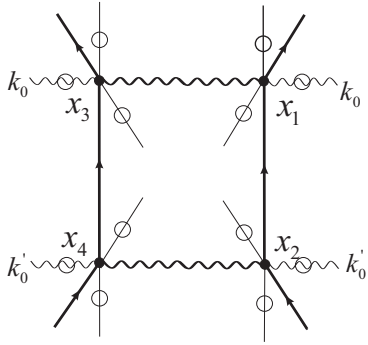}
\caption{Feynman diagram of a fourth-order S-matrix element of QED
perturbation theory which can be considered as an interpret of
Casimir effect. Lines with arrow denote to electric charged
particles overlapped by their vacuum-particles and lines with a
circle denote to their vacuum-particles. Wave lines denote to
photons overlapped by their vacuum-photons and wave lines with
circles denote to vacuum-photons.}
\end{figure}

Figure 4 is the Feynman diagram of a third-order S-matrix element
which can be described as the basic QED process of atomic
spontaneous emission. An electron is perturbed by the vacuum states
of electromagnetic field (vacuum-photons) at the space-time point
$x_2$ and excites a vacuum-photon with four-momentum $\frac{k}{2}$
into a photon with four-momentum $\frac{3k}{2}$ at $x_2$. Then the
electron becomes a off-shell electron and propagates to $x_1$ to
transfer the Coulomb interaction with a quark in the nuclear. After
that process the off-shell electron recovers to be on-shell at
$x_1$. Through the whole process, the electron transits from a
higher energy level at $x_2$ to a lower energy level at $x_1$ with
one photon emitted at $x_2$. The four-momentum of the emitted
one-photon state $\frac{3k}{2}$ is determine by the four-momentum of
the vacuum-photon $\frac{k}{2}$ and the photon $k$ at $x_2$.

\begin{figure}
\includegraphics[width=2.in]{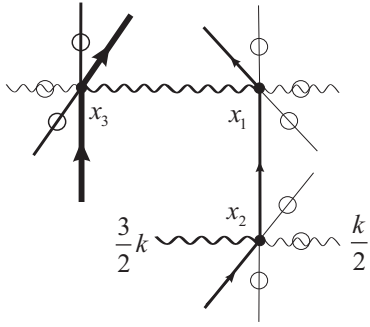}
\caption{Feynman diagram of a third-order S-matrix element of QED
perturbation theory which describes the basic process of atomic
spontaneous emission. Thick line with arrow denote to a quark
overlapped by its vacuum-particle, and thin line denote to an
electron overlapped by its vacuum-particle. Wave lines denote to
overlapped by their vacuum-photons and wave lines with circles
denote to vacuum-photons.}
\end{figure}

The above two physical phenomena has been widely studied in the
subject of quantum optics, where the vacuum state of electromagnetic
field is treated as plane waves which carries the vacuum energy of
the electromagnetic field. Because the possible particle-like
property of these plane waves is not concerned before, the above two
physical phenomena can not be studied by QED. After we considered
the particle-like property of these plane waves by introduce the
VPH, the interprets of the above two phenomena emerge from the QED
perturbation theory automatically. This result shows the existence
of the particle-like property of vacuum states is reasonable.

\section{discussions}

We have introduced the VPH to the QED perturbation theory. Because
the VPH is postulated for all kinds of quantum fields in the QFT,
there are vacuum-particles that correspond to quarks, leptons
besides electrons and positrons, and bosons besides photons. All of
them can be defined from the vacuum-momentum and vacuum-energy of
their fields like Eq.\ref{two} and Eq.\ref{three}. So the VPH can be
introduced into all non-abelian quantum gauge theories such as
electro-weak theory and quantum chromodynamics (QCD) without disturb
their calculation process just like in QED. This topic will be
interesting for the future study.

Besides, there is another important background theory of QFT in the
standard model which may be highly related to the vacuum-particles.
It is the spontaneous symmetry breaking. Since the Goldstone's
bosons emerges from the spontaneous continuous symmetry breaking
\cite{goldstone}, vacuum-particles of them can be easily defined.
However, Goldstone's bosons can not be observed in the nature
because the spontaneous breaking of gauge symmetry do not generate
them but generate Higgs bosons instead via Higgs mechanism
\cite{higgs}. It is necessary and interesting to discuss the Higgs
mechanism from the point of the vacuum-particle view. As we know,
before spontaneous symmetry breaking, the Higgs field is a complex
scalar field $\phi(x)$ with a non-zero vacuum expectation value,
which means its vacuum-particle does not have the lowest possible
energy and momentum. After the spontaneous symmetry breaking, a real
vector field $A_{\mu}(x)$ obtain the proper mass from the Higgs
field and the Higgs field becomes a Klein-Gorden field $\sigma(x)$.
This process can be interpreted as the vacuum-particles of the
vector field $A_{\mu}(x)$ obtain mass via coupling with the
vacuum-particles of the Higgs field $\phi(x)$, and at the mean time
the vacuum-particle of $\phi(x)$ becomes the vacuum-particle of
field $\sigma(x)$, whose one-particle excited state is a one Higgs
boson state. Thus in the VPH's point of view, the Higgs mechanism is
a process that happens due to the interaction between the
vacuum-particles of different fields.

In conclusion, the vacuum-particle hypothesis (VPH) can be obtained
naturally from the proposed particle-like property of vacuum state
of quantum fields by the wave-particle duality. It recovers the
energy and momentum of vacuum states, which benefit the
self-consistency of the quantum field theory (QFT) without
disturbing its mathematical form and calculation process. With the
VPH, the Casimir effect and atomic spontaneous emission which are
caused by the interaction between the matter and vacuum-state of the
electromagnetic field can be interpreted and calculated within the
frame of the QED perturbation theory directly. The Higgs mechanism
can also be interpreted in a more clearly physical picture by the
VPH.

\section*{Acknowledgment}

I thank my Ph.D supervisor Prof. Yu-zhu Wang for discussions and
valuable advices.

\end{document}